\documentclass[amsmath,amssymb,aps,prb,reprint,superscriptaddress]{revtex4-1}

\usepackage{comment}
\usepackage{chemformula} 
\usepackage[T1]{fontenc} 
\usepackage{graphicx}
\usepackage{dcolumn}
\usepackage{lipsum}
\usepackage{tabu} 
\usepackage{mathtools}

\usepackage{braket}
\usepackage{color}
\usepackage[11pt]{moresize}
\usepackage{anyfontsize}
\usepackage{bbding}
\usepackage{amsmath}
\usepackage{xcolor}
\usepackage{braket}
\usepackage[normalem]{ulem}

\begin{document} 

\author{Andrew N. Wakileh}
\affiliation{National Research Council of Canada, Ottawa, Ontario, Canada, K1A 0R6.}
\affiliation{Centre for Nanophotonics, Department of Physics, Engineering Physics, and Astronomy, Queen's University, Kingston, Ontario, Canada, K7L 3N6}
\author{Lingxi Yu}
\affiliation{National Research Council of Canada, Ottawa, Ontario, Canada, K1A 0R6.}
\affiliation{Department of Physics, University of Ottawa, Ottawa, Ontario, Canada, K1N 6N5}
\author{Do\u{g}a Dokuz}
\affiliation{National Research Council of Canada, Ottawa, Ontario, Canada, K1A 0R6.}
\affiliation{Department of Physics, University of Ottawa, Ottawa, Ontario, Canada, K1N 6N5}
\author{Sofiane Haffouz}
\affiliation{National Research Council of Canada, Ottawa, Ontario, Canada, K1A 0R6.}
\author{Xiaohua Wu}
\affiliation{National Research Council of Canada, Ottawa, Ontario, Canada, K1A 0R6.}
\author{Jean Lapointe}
\affiliation{National Research Council of Canada, Ottawa, Ontario, Canada, K1A 0R6.}
\author{David B. Northeast}
\affiliation{National Research Council of Canada, Ottawa, Ontario, Canada, K1A 0R6.}
\author{Robin L. Williams}
\affiliation{National Research Council of Canada, Ottawa, Ontario, Canada, K1A 0R6.}
\author{Nir Rotenberg}
\affiliation{Centre for Nanophotonics, Department of Physics, Engineering Physics, and Astronomy, Queen's University, Kingston, Ontario, Canada, K7L 3N6}
\author{Philip J. Poole}
\affiliation{National Research Council of Canada, Ottawa, Ontario, Canada, K1A 0R6.}
\author{Dan Dalacu}
\affiliation{National Research Council of Canada, Ottawa, Ontario, Canada, K1A 0R6.}
\affiliation{Centre for Nanophotonics, Department of Physics, Engineering Physics, and Astronomy, Queen's University, Kingston, Ontario, Canada, K7L 3N6}
\affiliation{Department of Physics, University of Ottawa, Ottawa, Ontario, Canada, K1N 6N5}
\email{dan.dalacu@nrc.ca}

\title{On-demand single photon emission in the telecom C-band from nanowire-based quantum dots}




\begin{abstract}

Single photon sources operating on-demand at telecom wavelengths are required in fiber-based quantum secure communication technologies. In this work we demonstrate single photon emission from position-controlled nanowire quantum dots emitting at $\lambda>1530$\,nm. Using above-band pulsed excitation, we obtain single photon purities of $g^{(2)}(0) = 0.062$. These results represent an important step towards the scalable manufacture of high efficiency, high rate single photon emitters in the telecom C-band.

\end{abstract}

\maketitle 

Long-haul terrestrial optical fiber networks operate at wavelengths in the C-band ($\lambda=1530-1565$\,nm) where transmission losses are minimum. In quantum secure communications, photon loss is particularly important as it presents a loophole allowing eavesdroppers to go undetected. As such, there is currently a strong research effort to develop single photon sources operating at high rates and high efficiencies in the telecom C-band\cite{Cao_JOS2019}.  

Solid-state single photon emitters, and in particular, semiconductor quantum dot emitters\cite{Arakawa_APR2020}, are a promising candidate: they offer on-demand operation at high rates and can be incorporated in photonic structures that allow for high efficiency collection. To date, quantum dots grown in both the InAs/InP\cite{Miyazawa_JJAP2005,Takemoto_JAP2007,Takemoto_APE2010,Birowosuto_SR2012,Miyazawa_APL2016,Muller_NC2018,Musial_AQT2020,Anderson_QI2020,Musial_APL2021,Anderson_APL2021,Holewa_ACSP2022} and InGaAs/GaAs\cite{Paul_APL2017,Carmesin_2018,Nawrath_APL2019,Bauer_APL2021,Nawrath_APL2021,Zeuner_ACSP2021,Dusanowski_NC2022} material systems have demonstrated single photon emission in the C-band, including on-demand generation\cite{Takemoto_SR2015,Miyazawa_APL2016,Paul_APL2017,Musial_AQT2020,Nawrath_APL2021}. Current efforts are focused on establishing the indistinguishability of sequential photons emitted from such sources using two-photon interference measurements \cite{Nawrath_APL2019,Nawrath_APL2021,Anderson_APL2021}. The above platforms all employ quantum dots grown via Stranski-Krastanov nucleation or droplet epitaxy\cite{Anderson_APL2021}, meaning that the dots nucleate at random positions on the substrate and are, thus, difficult to scalably incorporate into devices.

In contrast, quantum dots grown by selective-area vapour-liquid-solid epitaxy (SA-VLS)\cite{Dalacu_APL2011} are well-defined sections of a lower bandgap semiconductor surrounded by a higher bandgap nanowire. Such nanowire-based quantum dots can be incorporated in photonic structures that provide high efficiency collection\cite{Dalacu_NT2019} and importantly, are by definition grown with position-control, offering much desired scalability\cite{Laferriere_SR2022}. To date, we have demonstrated single photon emission from nanowire quantum dots from wavelengths $\lambda < 1\,\mu$m\cite{Dalacu_NL2012} to the telecom O-band\cite{Laferriere_NL2023}.


In this article, we extend the operating range of these site-selected single photon sources to the telecom C-band. Using above-band, pulsed excitation we obtain single photon emission at $\lambda = 1531$\,nm with a probability of multiphoton emission of $g^{(2)}(0) = 0.062$. These results open possibilities for generating single photons across the C-band using structures fabricated with a scalable approach that is compatible with the silicon material system\cite{Chauvin_APL2012}.

\begin{figure}
\begin{center}
\includegraphics*[width=7.5cm,clip=true]{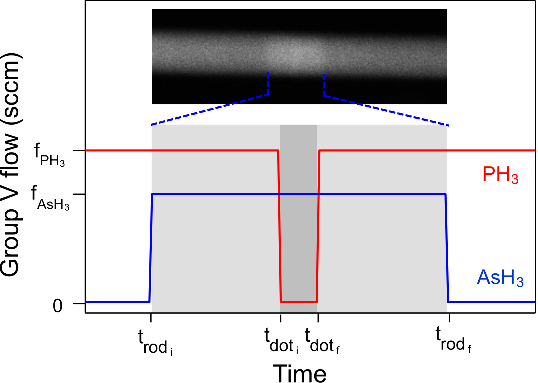}
\end{center}
\caption{Group V switching sequence during growth of the dot-in-a-rod section of the nanowire core. Light (dark) grey region indicates rod (dot) growth. Inset shows a transmission electron microscopy image of the corresponding section.}\label{Fig_1}
\end{figure}

The nanowire quantum dots used in this work are based on InAs$_x$P$_{1-x}$ sections in bottom-up InP nanowire cores grown using S-A VLS epitaxy, described in detail elsewhere \cite{Dalacu_APL2011,Laferriere_SR2022}. Such structures typically operate at $\lambda \sim 950$\,nm due to a low arsenic concentration of $x\sim 25\%$.  To shift to longer wavelengths, we adopt a dot-in-a-rod structure \cite{Haffouz_APL2020} previously used to demonstrate efficient single photon generation in the O-band\cite{Laferriere_NL2023}. In this structure, the dot is incorporated within an InAs$_y$P$_{1-y}$ rod which is embedded within the InP nanowire core. To grow these structures, we employ the switching sequence shown in Fig.~\ref{Fig_1}. The rod is defined by introducing AsH$_3$ into the growth chamber for a time $t_\mathrm{rod}=t_\mathrm{rod_f}-t_\mathrm{rod_i}$ whilst the dot is defined by switching off the PH$_3$ flow for $t_\mathrm{dot}=t_\mathrm{dot_f}-t_\mathrm{dot_i}$.


Four samples were grown, labelled  A, B, C and D, where the growth conditions for the  nanowire core were varied as shown in Table~\ref{Table_1}. Sample A was grown using the same conditions previously used to obtain O-band emission, namely the rod was grown using PH$_3$ and AsH$_3$ flow rates of $f_\mathrm{PH_3}=2$\,sccm and $f_\mathrm{AsH_3}=1$\,sccm, respectively. In Samples B and C, the AsH$_3$ flow rate was increased to $f_\mathrm{PH_3}=1.5$\,sccm and $f_\mathrm{PH_3}=2$\,sccm, respectively, whilst in D, the AsH$_3$ flow rate was the same as in sample B but the dot growth time was increased to 6.5 seconds. In all cases the nanowire cores were clad with an InP shell to increase the total diameter of the nanowire. This diameter, $D$, is an important parameter in designing the nanowire structures: it determines the radiative recombination lifetime of excitons in the quantum dot and for insufficiently clad nanowire cores may result in inhibited spontaneous emission rates\cite{Haffouz_NL2018}. Here we target diameters $D=\lambda/4$ for which $D=387.5$\,nm at an emission wavelength of $\lambda=1550$\,nm.

\begin{table} 
\caption{Group V flow rates, $f_\mathrm{PH_3}$ and $f_\mathrm{AsH_3}$, used during rod growth and rod (dot) growth times, $t_\mathrm{rod}$ ($t_\mathrm{dot}$). \label{Table_1}}
\setlength{\tabcolsep}{12.pt}
\begin{tabular}{ccccc}
\hline
Sample & $f_\mathrm{PH_3}$ & $f_\mathrm{AsH_3}$  & $t_\mathrm{rod}$ & $t_\mathrm{dot}$ \\
& (sccm) & (sccm) & (s) & (s) \\
\hline
A   &   2 &1    &  20 & 3.5    \\ 
B   &   2  & 1.5     &  20 & 3.5    \\ 
C   &   2  & 2     &  20 & 3.5     \\  
D   &   2  & 1.5     &  20 & 6.5  \\  
\hline
\end{tabular}
\end{table}

\begin{figure}[htb]
\begin{center}
\includegraphics*[width=7.cm,clip=true]{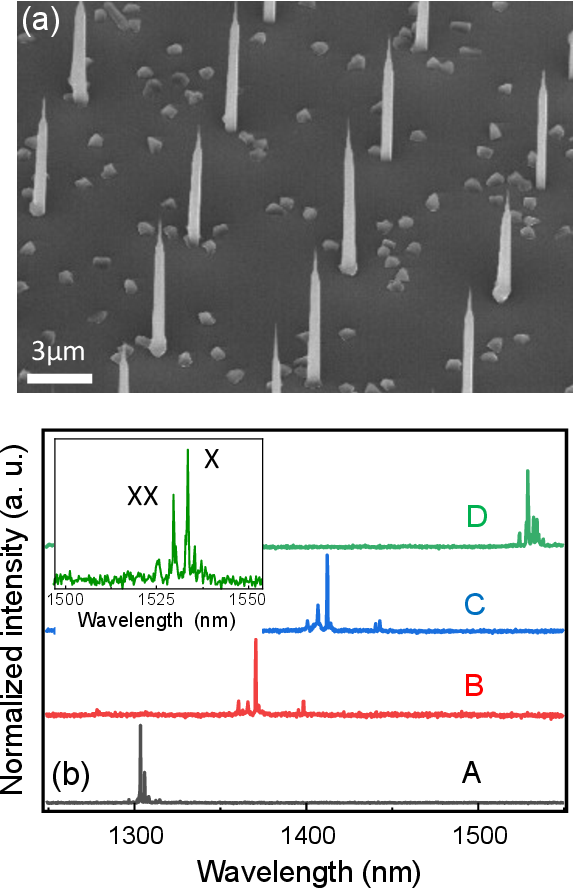}
\end{center}
\caption{(a) Scanning electron micrograph imaged at 45$^\circ$ of an array of nanowires corresponding to Sample D. (b) Photoluminescence spectra from each sample measured at excitation powers close to saturation. Inset shows the emission from Sample D at low excitation over a reduced spectral range.}\label{Fig_2}
\end{figure}

Spectrally-resolved photoluminescence (PL) measurements at 4\,K were made in a closed-cycle He cryostat. Individual nanowires from a patterned array were excited above-band using continuous wave (CW) excitation at $\lambda = 780$\,nm through a 100x cryogenic objective (numerical aperture = 0.81), see Fig.~\ref{Fig_2}(a). The PL was collected through the same objective, coupled into a SMF-28 fiber and directed to a spectrometer equipped with a liquid nitrogen-cooled InGaAs linear array detector. Exemplary PL spectra of individual nanowire quantum dots from each sample, measured at excitation powers close to saturation, are shown in Fig.~\ref{Fig_2}(b). Each spectrum consists of several discrete peaks, each peak corresponding to a different excitonic complex in the s-orbital of the dot. At low excitation power, we typically observe two dominant peaks (see inset) which we associate with emission from the neutral exciton, $X$, and the biexciton $XX$. 

The s-shell emission from Sample A is observed at $\lambda \sim 1300$\,nm, as expected from the previous study\cite{Laferriere_NL2023}. Increasing the AsH$_3$ flow to 1.5\,sccm (Sample B) results in a redshift of 70\,nm. This shift, due to a reduction in confinement from the barrier and in the dot, is a consequence of the higher arsenic concentration in the rod and dot, respectively. We note that incorporation of the rod will also increase the arsenic concentration in the dot for a given AsH$_3$ flow by reducing memory effects in the reaction chamber i.e. the occurance of phosphorous tailing.

A further increase of the AsH$_3$ flow to 2\,sccm (Sample C) produces an additional redshift of 40\,nm for an emission wavelength of $\lambda \sim 1410$\,nm. If, instead of reducing the confinement using composition, we simply increase the thickness of the dot (Sample D), we obtain a much larger redshift of 160\,nm, (compare with Sample B) such that this sample emits at $\lambda > 1530$\,nm (i.e. in the C-band).  

For this sample, we determined the diameter of the nanowire from scanning electron microscopy images, Fig.~\ref{Fig_2}(a). We measured values of $D\sim 490$\,nm which correspond to $D/\lambda ~\sim 0.32$ at the emission wavelength $\lambda = 1530$\,nm i.e. significantly larger than the optimal value of $D/\lambda=0.23$ \cite{Dalacu_NT2019}.  For devices with such a high $D/\lambda$ ratio, there will be a reduction of the total emission rate in addition to the coupling into the mode of interest, HE$_{11}$, of nearly a factor of two in both cases. We also note that, unlike the optimized devices, this structure supports higher order TE$_{01}$ and TM$_{01}$ modes, although, this should not impact performance since an emitter on-axis of an ideal nanowire does not couple to these modes.

\begin{figure}
   \includegraphics*[width=8.cm,clip=true]{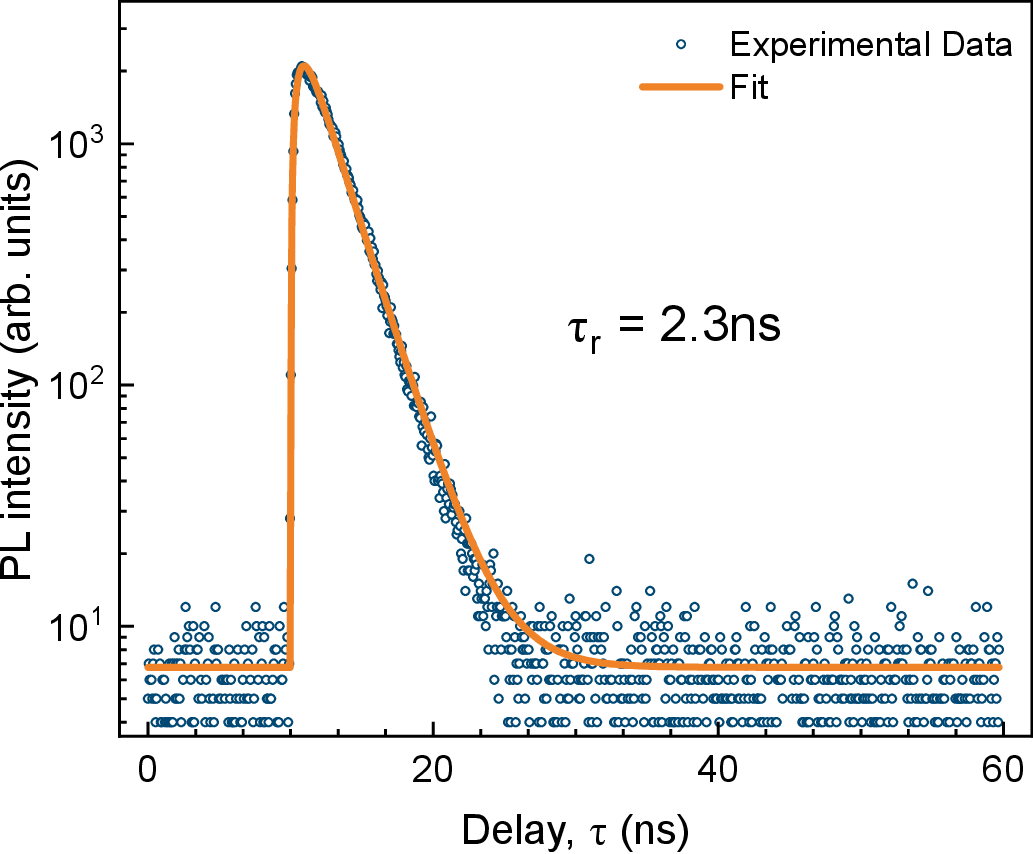}
    \caption{Time-resolved photoluminescence of the $XX$ peak emitting at $\lambda=1531$\,nm. Orange curve is an exponential fit described in the text.}
    \label{Fig_3}
\end{figure}

To evaluate the impact of the high $D/\lambda$ ratio on the emission rate, we measured the radiative lifetime of the $XX$ peak. The peak was selected using a tunable filter (bandwidth $0.1-3\,$pm) and sent to a fiber-coupled superconducting nanowire single photon detector (SNSPD). Excitation was above-band ($\lambda = 670$\,nm) with a 100\,ps pulsed laser at a repetition rate of 5\,MHz. The PL decay curve, plotted in Fig.~\ref{Fig_3}, shows a simple mono-exponential behaviour which we fit with an expression of the form $\propto [1-e^{-(\tau-\tau_{d})/\tau_{e}}]e^{-(\tau-\tau_{d})/\tau_{r}}$ where $\tau_{d}$ is an offset delay and $\tau_{e}$ ($\tau_{r}$) is the biexciton PL rise (decay) time. From the fit, we extract a radiative lifetime $\tau_{r} = 2.3$ \,ns. This value is $\sim 2\times$ larger than that typically observed for the $XX$ decay in optimized nanowire structures, consistent with the expected increase in lifetime based on the $D/\lambda$ ratio\cite{XX}.

\begin{figure}[htb]
   \includegraphics*[width=8cm,clip=true]{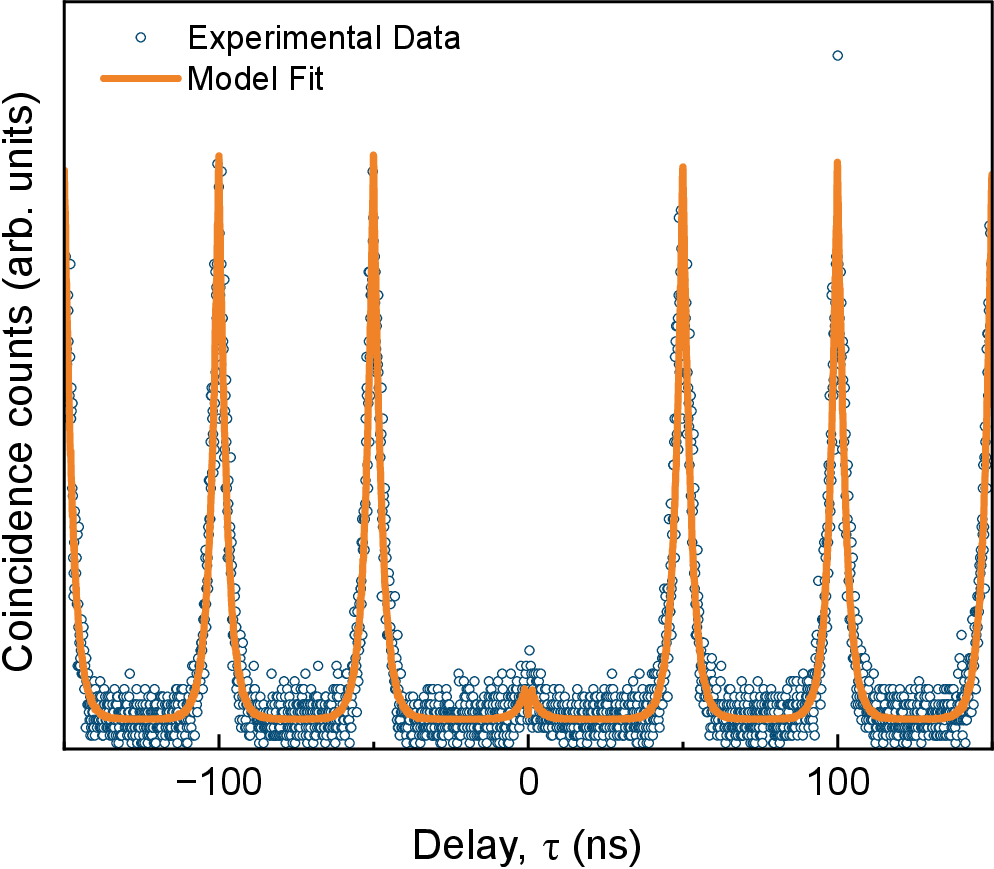}
    \caption{Second-order auto-correlation measurement, $g^{(2)}(\tau)$, of the $XX$ peak at $\lambda=1531$\,nm from a nanowire in Sample D. The fit (orange curve) is generated using a stochastic model\cite{Mnaymneh_AQT2020}.}
    \label{Fig_4}
\end{figure}

To verify the single photon purity of $XX$ peak at $\lambda =1531$\,nm we performed a second-order correlation measurement using a fiber-based Hanbury–Brown-Twiss setup. Here, the filtered $XX$ emission was sent to two SNSPDs via a 50:50 fiber beamsplitter and excitation was with the $\lambda = 670$\,nm pulsed source but with the repetition rate increased to 20\,MHz. The measured coincidences, $g^{(2)}(\tau)$, are plotted in Fig.~\ref{Fig_4} (blue open circles) and show a significant reduction of the peak at $\tau = 0$ relative to the side peaks indicative of high purity single photon emission. 

The measured correlations were simulated using a stochastic model described in detail elsewhere\cite{Laferriere_NL2020,Mnaymneh_AQT2020}. In the simulation, coincidences around zero delay are attributed to a re-excitation process\cite{Aichele_NJP2004,Santori_NJP2004} and modelled as a competition between the band-to-band decay rate in the barrier, the carrier capture rate into the dot, and the radiative decay rate of the exciton. We assume negligible contribution from uncorrelated emission (e.g. spectral pollution from nearby emitters) since the nanowire device contains a single quantum dot. Using the model fit (orange curve in Fig.~\ref{Fig_4}) we calculate the single photon purity from the ratio of coincidences in the zero-delay peak relative to the side peaks. We obtain a raw value of $g^{(2)}(0) = 0.097$ which is reduced to $g^{(2)}(0) = 0.062$ after background subtraction.

Finally, we note that the count rates of the C-band emitter were significantly reduced, by a factor of $\sim 50 \times$, compared to an optimized O-band device\cite{Laferriere_NL2023}. This reduction was observed regardless of the type of excitation i.e. CW versus pulsed, suggesting that device efficiency rather than the transition lifetime is responsible. The lower efficiencies can result from less efficient coupling of the emitter to the fundamental waveguide mode, $\beta_{\mathrm{HE_{11}}}$, due to the high $D/\lambda$ ratio, as discussed above. Further, since we have increased the dot thickness to achieve C-band emission, there may be mixing of the valence ground state resulting in excitons with polarizations that do not couple to the HE$_\mathrm{11}$\cite{Haffouz_NL2018}. Lastly, the taper geometry of the nanowire may not be optimal for C-band emission\cite{Dalacu_NANOM2021} resulting in a reduced collection efficiency of the HE$_\mathrm{11}$ into the external optical system. We speculate that minor modifications in the nanowire design, implemented in future growths, will produce devices with a greater than 10-fold improvement in efficiency. 

In summary, we have extended our previous work on nanowire quantum dot telecom emitters to demonstrate, to our knowledge for the first time, C-band single photon emission from such structures. We measured single photon purities of $1-g^{(2)}(0)>0.9$ at an emission wavelength of $\lambda=1531$\,nm. Residual multi-photon emission events were attributed to re-excitation of the quantum dot from the same excitation pulse. With increased collection efficiencies and higher emission rates anticipated in optimized structures, position-controlled nanowires offer a viable route for the scalable manufacture of single photon emitters for fiber-based quantum secure communications technologies. 

This work was supported by the Natural Sciences and Engineering Research Council of Canada through the Discovery Grant SNQLS, the National Research Council of Canada through the Quantum Sensing Challenge Program project `Telecom Photonic Resources for Quantum Sensing' and the Canadian Space Agency through the collaborative project ‘Field Deployable Single Photon Emitters for Quantum Secured Communications'. 

\textbf{DATA AVAILABILITY}
The data that support the findings of this study are available from the corresponding author upon reasonable request.

\bibliography{whiskers}

\begin{thebibliography}{35}%
\makeatletter
\providecommand \@ifxundefined [1]{%
 \@ifx{#1\undefined}
}%
\providecommand \@ifnum [1]{%
 \ifnum #1\expandafter \@firstoftwo
 \else \expandafter \@secondoftwo
 \fi
}%
\providecommand \@ifx [1]{%
 \ifx #1\expandafter \@firstoftwo
 \else \expandafter \@secondoftwo
 \fi
}%
\providecommand \natexlab [1]{#1}%
\providecommand \enquote  [1]{``#1''}%
\providecommand \bibnamefont  [1]{#1}%
\providecommand \bibfnamefont [1]{#1}%
\providecommand \citenamefont [1]{#1}%
\providecommand \href@noop [0]{\@secondoftwo}%
\providecommand \href [0]{\begingroup \@sanitize@url \@href}%
\providecommand \@href[1]{\@@startlink{#1}\@@href}%
\providecommand \@@href[1]{\endgroup#1\@@endlink}%
\providecommand \@sanitize@url [0]{\catcode `\\12\catcode `\$12\catcode
  `\&12\catcode `\#12\catcode `\^12\catcode `\_12\catcode `\%12\relax}%
\providecommand \@@startlink[1]{}%
\providecommand \@@endlink[0]{}%
\providecommand \url  [0]{\begingroup\@sanitize@url \@url }%
\providecommand \@url [1]{\endgroup\@href {#1}{\urlprefix }}%
\providecommand \urlprefix  [0]{URL }%
\providecommand \Eprint [0]{\href }%
\providecommand \doibase [0]{http://dx.doi.org/}%
\providecommand \selectlanguage [0]{\@gobble}%
\providecommand \bibinfo  [0]{\@secondoftwo}%
\providecommand \bibfield  [0]{\@secondoftwo}%
\providecommand \translation [1]{[#1]}%
\providecommand \BibitemOpen [0]{}%
\providecommand \bibitemStop [0]{}%
\providecommand \bibitemNoStop [0]{.\EOS\space}%
\providecommand \EOS [0]{\spacefactor3000\relax}%
\providecommand \BibitemShut  [1]{\csname bibitem#1\endcsname}%
\let\auto@bib@innerbib\@empty
\bibitem [{\citenamefont {X.~Cao}(2019)}]{Cao_JOS2019}%
  \BibitemOpen
  \bibfield  {author} {\bibinfo {author} {\bibfnamefont {F.~D.}\ \bibnamefont
  {X.~Cao}, \bibfnamefont {M.~Zopf}},\ }\href@noop {} {\bibfield  {journal}
  {\bibinfo  {journal} {J. Semicond.}\ }\textbf {\bibinfo {volume} {40}},\
  \bibinfo {pages} {071901} (\bibinfo {year} {2019})}\BibitemShut {NoStop}%
\bibitem [{\citenamefont {Arakawa}\ and\ \citenamefont
  {Holmes}(2020)}]{Arakawa_APR2020}%
  \BibitemOpen
  \bibfield  {author} {\bibinfo {author} {\bibfnamefont {Y.}~\bibnamefont
  {Arakawa}}\ and\ \bibinfo {author} {\bibfnamefont {M.~J.}\ \bibnamefont
  {Holmes}},\ }\href@noop {} {\bibfield  {journal} {\bibinfo  {journal} {Appl.
  Phys. Rev.}\ }\textbf {\bibinfo {volume} {7}},\ \bibinfo {pages} {021309}
  (\bibinfo {year} {2020})}\BibitemShut {NoStop}%
\bibitem [{\citenamefont {Miyazawa}\ \emph {et~al.}(2005)\citenamefont
  {Miyazawa}, \citenamefont {Takemoto}, \citenamefont {Sakuma}, \citenamefont
  {Hirose}, \citenamefont {Usuki}, \citenamefont {Yokoyama}, \citenamefont
  {Takatsu},\ and\ \citenamefont {Arakawa}}]{Miyazawa_JJAP2005}%
  \BibitemOpen
  \bibfield  {author} {\bibinfo {author} {\bibfnamefont {T.}~\bibnamefont
  {Miyazawa}}, \bibinfo {author} {\bibfnamefont {K.}~\bibnamefont {Takemoto}},
  \bibinfo {author} {\bibfnamefont {Y.}~\bibnamefont {Sakuma}}, \bibinfo
  {author} {\bibfnamefont {S.}~\bibnamefont {Hirose}}, \bibinfo {author}
  {\bibfnamefont {T.}~\bibnamefont {Usuki}}, \bibinfo {author} {\bibfnamefont
  {N.}~\bibnamefont {Yokoyama}}, \bibinfo {author} {\bibfnamefont
  {M.}~\bibnamefont {Takatsu}}, \ and\ \bibinfo {author} {\bibfnamefont
  {Y.}~\bibnamefont {Arakawa}},\ }\href@noop {} {\bibfield  {journal} {\bibinfo
   {journal} {J. J. Appl. Phys.}\ }\textbf {\bibinfo {volume} {44}},\ \bibinfo
  {pages} {L620} (\bibinfo {year} {2005})}\BibitemShut {NoStop}%
\bibitem [{\citenamefont {Takemoto}\ \emph {et~al.}(2007)\citenamefont
  {Takemoto}, \citenamefont {Takatsu}, \citenamefont {Hirose}, \citenamefont
  {Yokoyama}, \citenamefont {Sakuma}, \citenamefont {Usuki}, \citenamefont
  {Miyazawa},\ and\ \citenamefont {Arakawa}}]{Takemoto_JAP2007}%
  \BibitemOpen
  \bibfield  {author} {\bibinfo {author} {\bibfnamefont {K.}~\bibnamefont
  {Takemoto}}, \bibinfo {author} {\bibfnamefont {M.}~\bibnamefont {Takatsu}},
  \bibinfo {author} {\bibfnamefont {S.}~\bibnamefont {Hirose}}, \bibinfo
  {author} {\bibfnamefont {N.}~\bibnamefont {Yokoyama}}, \bibinfo {author}
  {\bibfnamefont {Y.}~\bibnamefont {Sakuma}}, \bibinfo {author} {\bibfnamefont
  {T.}~\bibnamefont {Usuki}}, \bibinfo {author} {\bibfnamefont
  {T.}~\bibnamefont {Miyazawa}}, \ and\ \bibinfo {author} {\bibfnamefont
  {Y.}~\bibnamefont {Arakawa}},\ }\href@noop {} {\bibfield  {journal} {\bibinfo
   {journal} {J. Appl. Phys.}\ }\textbf {\bibinfo {volume} {101}},\ \bibinfo
  {pages} {081720} (\bibinfo {year} {2007})}\BibitemShut {NoStop}%
\bibitem [{\citenamefont {Takemoto}\ \emph {et~al.}(2010)\citenamefont
  {Takemoto}, \citenamefont {Nambu}, \citenamefont {Miyazawa}, \citenamefont
  {Wakui}, \citenamefont {Hirose}, \citenamefont {Usuki}, \citenamefont
  {Takutsu}, \citenamefont {Yokoyama}, \citenamefont {Yoshino},\ and\
  \citenamefont {Tomita}}]{Takemoto_APE2010}%
  \BibitemOpen
  \bibfield  {author} {\bibinfo {author} {\bibfnamefont {K.}~\bibnamefont
  {Takemoto}}, \bibinfo {author} {\bibfnamefont {Y.}~\bibnamefont {Nambu}},
  \bibinfo {author} {\bibfnamefont {T.}~\bibnamefont {Miyazawa}}, \bibinfo
  {author} {\bibfnamefont {K.}~\bibnamefont {Wakui}}, \bibinfo {author}
  {\bibfnamefont {S.}~\bibnamefont {Hirose}}, \bibinfo {author} {\bibfnamefont
  {T.}~\bibnamefont {Usuki}}, \bibinfo {author} {\bibfnamefont
  {M.}~\bibnamefont {Takutsu}}, \bibinfo {author} {\bibfnamefont
  {N.}~\bibnamefont {Yokoyama}}, \bibinfo {author} {\bibfnamefont
  {K.}~\bibnamefont {Yoshino}}, \ and\ \bibinfo {author} {\bibfnamefont
  {A.}~\bibnamefont {Tomita}},\ }\href@noop {} {\bibfield  {journal} {\bibinfo
  {journal} {Appl. Phys. Exp.}\ }\textbf {\bibinfo {volume} {3}},\ \bibinfo
  {pages} {092802} (\bibinfo {year} {2010})}\BibitemShut {NoStop}%
\bibitem [{\citenamefont {Birowosuto}\ \emph {et~al.}(2012)\citenamefont
  {Birowosuto}, \citenamefont {Sumikura}, \citenamefont {Matsuo}, \citenamefont
  {Taniyama}, \citenamefont {van Veldhoven}, \citenamefont {N{\"o}tzel},\ and\
  \citenamefont {Notomi}}]{Birowosuto_SR2012}%
  \BibitemOpen
  \bibfield  {author} {\bibinfo {author} {\bibfnamefont {M.}~\bibnamefont
  {Birowosuto}}, \bibinfo {author} {\bibfnamefont {H.}~\bibnamefont
  {Sumikura}}, \bibinfo {author} {\bibfnamefont {S.}~\bibnamefont {Matsuo}},
  \bibinfo {author} {\bibfnamefont {H.}~\bibnamefont {Taniyama}}, \bibinfo
  {author} {\bibfnamefont {P.}~\bibnamefont {van Veldhoven}}, \bibinfo {author}
  {\bibfnamefont {R.}~\bibnamefont {N{\"o}tzel}}, \ and\ \bibinfo {author}
  {\bibfnamefont {M.}~\bibnamefont {Notomi}},\ }\href@noop {} {\bibfield
  {journal} {\bibinfo  {journal} {Sci. Rep.}\ }\textbf {\bibinfo {volume}
  {2}},\ \bibinfo {pages} {321} (\bibinfo {year} {2012})}\BibitemShut {NoStop}%
\bibitem [{\citenamefont {Miyazawa}\ \emph {et~al.}(2016)\citenamefont
  {Miyazawa}, \citenamefont {Takemoto}, \citenamefont {Nambu}, \citenamefont
  {Miki}, \citenamefont {Yamashita}, \citenamefont {Terai}, \citenamefont
  {Fujiwara}, \citenamefont {Sasaki}, \citenamefont {Sakuma}, \citenamefont
  {Takatsu}, \citenamefont {Yamamoto},\ and\ \citenamefont
  {Arakawa}}]{Miyazawa_APL2016}%
  \BibitemOpen
  \bibfield  {author} {\bibinfo {author} {\bibfnamefont {T.}~\bibnamefont
  {Miyazawa}}, \bibinfo {author} {\bibfnamefont {K.}~\bibnamefont {Takemoto}},
  \bibinfo {author} {\bibfnamefont {Y.}~\bibnamefont {Nambu}}, \bibinfo
  {author} {\bibfnamefont {S.}~\bibnamefont {Miki}}, \bibinfo {author}
  {\bibfnamefont {T.}~\bibnamefont {Yamashita}}, \bibinfo {author}
  {\bibfnamefont {H.}~\bibnamefont {Terai}}, \bibinfo {author} {\bibfnamefont
  {M.}~\bibnamefont {Fujiwara}}, \bibinfo {author} {\bibfnamefont
  {M.}~\bibnamefont {Sasaki}}, \bibinfo {author} {\bibfnamefont
  {Y.}~\bibnamefont {Sakuma}}, \bibinfo {author} {\bibfnamefont
  {M.}~\bibnamefont {Takatsu}}, \bibinfo {author} {\bibfnamefont
  {T.}~\bibnamefont {Yamamoto}}, \ and\ \bibinfo {author} {\bibfnamefont
  {Y.}~\bibnamefont {Arakawa}},\ }\href@noop {} {\bibfield  {journal} {\bibinfo
   {journal} {Appl. Phys. Lett.}\ }\textbf {\bibinfo {volume} {109}},\ \bibinfo
  {pages} {132106} (\bibinfo {year} {2016})}\BibitemShut {NoStop}%
\bibitem [{\citenamefont {M{\"u}ller}\ \emph {et~al.}(2018)\citenamefont
  {M{\"u}ller}, \citenamefont {Skiba-Szymanska}, \citenamefont {Krysa},
  \citenamefont {Huwer}, \citenamefont {Felle}, \citenamefont {Anderson},
  \citenamefont {Stevenson}, \citenamefont {Heffernan}, \citenamefont
  {Ritchie},\ and\ \citenamefont {Shields}}]{Muller_NC2018}%
  \BibitemOpen
  \bibfield  {author} {\bibinfo {author} {\bibfnamefont {T.}~\bibnamefont
  {M{\"u}ller}}, \bibinfo {author} {\bibfnamefont {J.}~\bibnamefont
  {Skiba-Szymanska}}, \bibinfo {author} {\bibfnamefont {A.}~\bibnamefont
  {Krysa}}, \bibinfo {author} {\bibfnamefont {J.}~\bibnamefont {Huwer}},
  \bibinfo {author} {\bibfnamefont {M.}~\bibnamefont {Felle}}, \bibinfo
  {author} {\bibfnamefont {M.}~\bibnamefont {Anderson}}, \bibinfo {author}
  {\bibfnamefont {R.}~\bibnamefont {Stevenson}}, \bibinfo {author}
  {\bibfnamefont {J.}~\bibnamefont {Heffernan}}, \bibinfo {author}
  {\bibfnamefont {D.}~\bibnamefont {Ritchie}}, \ and\ \bibinfo {author}
  {\bibfnamefont {A.}~\bibnamefont {Shields}},\ }\href@noop {} {\bibfield
  {journal} {\bibinfo  {journal} {Nat. Commun.}\ }\textbf {\bibinfo {volume}
  {9}},\ \bibinfo {pages} {862} (\bibinfo {year} {2018})}\BibitemShut {NoStop}%
\bibitem [{\citenamefont {Musiał}\ \emph {et~al.}(2020)\citenamefont
  {Musiał}, \citenamefont {Holewa}, \citenamefont {Wyborski}, \citenamefont
  {Syperek}, \citenamefont {Kors}, \citenamefont {Reithmaier}, \citenamefont
  {Sek},\ and\ \citenamefont {Benyouce}}]{Musial_AQT2020}%
  \BibitemOpen
  \bibfield  {author} {\bibinfo {author} {\bibfnamefont {A.}~\bibnamefont
  {Musiał}}, \bibinfo {author} {\bibfnamefont {P.}~\bibnamefont {Holewa}},
  \bibinfo {author} {\bibfnamefont {P.}~\bibnamefont {Wyborski}}, \bibinfo
  {author} {\bibfnamefont {M.}~\bibnamefont {Syperek}}, \bibinfo {author}
  {\bibfnamefont {A.}~\bibnamefont {Kors}}, \bibinfo {author} {\bibfnamefont
  {J.~P.}\ \bibnamefont {Reithmaier}}, \bibinfo {author} {\bibfnamefont
  {G.}~\bibnamefont {Sek}}, \ and\ \bibinfo {author} {\bibfnamefont
  {M.}~\bibnamefont {Benyouce}},\ }\href@noop {} {\bibfield  {journal}
  {\bibinfo  {journal} {Adv. Quant. Technol.}\ }\textbf {\bibinfo {volume}
  {3}},\ \bibinfo {pages} {1900082} (\bibinfo {year} {2020})}\BibitemShut
  {NoStop}%
\bibitem [{\citenamefont {Anderson}\ \emph {et~al.}(2020)\citenamefont
  {Anderson}, \citenamefont {M{\"u}ller}, \citenamefont {Huwer}, \citenamefont
  {Skiba-Szymanska}, \citenamefont {Krysa}, \citenamefont {Stevenson},
  \citenamefont {Heffernan}, \citenamefont {Ritchie},\ and\ \citenamefont
  {Shields}}]{Anderson_QI2020}%
  \BibitemOpen
  \bibfield  {author} {\bibinfo {author} {\bibfnamefont {M.}~\bibnamefont
  {Anderson}}, \bibinfo {author} {\bibfnamefont {T.}~\bibnamefont
  {M{\"u}ller}}, \bibinfo {author} {\bibfnamefont {J.}~\bibnamefont {Huwer}},
  \bibinfo {author} {\bibfnamefont {J.}~\bibnamefont {Skiba-Szymanska}},
  \bibinfo {author} {\bibfnamefont {A.~B.}\ \bibnamefont {Krysa}}, \bibinfo
  {author} {\bibfnamefont {R.~M.}\ \bibnamefont {Stevenson}}, \bibinfo {author}
  {\bibfnamefont {J.}~\bibnamefont {Heffernan}}, \bibinfo {author}
  {\bibfnamefont {D.~A.}\ \bibnamefont {Ritchie}}, \ and\ \bibinfo {author}
  {\bibfnamefont {A.~J.}\ \bibnamefont {Shields}},\ }\href@noop {} {\bibfield
  {journal} {\bibinfo  {journal} {NPJ Quant. Inform.}\ }\textbf {\bibinfo
  {volume} {6}},\ \bibinfo {pages} {14} (\bibinfo {year} {2020})}\BibitemShut
  {NoStop}%
\bibitem [{\citenamefont {Musiał}\ \emph {et~al.}(2021)\citenamefont
  {Musiał}, \citenamefont {Mikulicz}, \citenamefont {Zieli{\'n}ska},
  \citenamefont {Sitarek}, \citenamefont {Wyborski}, \citenamefont {Kuniej},
  \citenamefont {Reithmaier}, \citenamefont {S{\c e}k},\ and\ \citenamefont
  {Benyoucef}}]{Musial_APL2021}%
  \BibitemOpen
  \bibfield  {author} {\bibinfo {author} {\bibfnamefont {A.}~\bibnamefont
  {Musiał}}, \bibinfo {author} {\bibfnamefont {P.}~\bibnamefont {Mikulicz}},
  \bibinfo {author} {\bibfnamefont {A.}~\bibnamefont {Zieli{\'n}ska}}, \bibinfo
  {author} {\bibfnamefont {P.}~\bibnamefont {Sitarek}}, \bibinfo {author}
  {\bibfnamefont {P.}~\bibnamefont {Wyborski}}, \bibinfo {author}
  {\bibfnamefont {M.}~\bibnamefont {Kuniej}}, \bibinfo {author} {\bibfnamefont
  {J.}~\bibnamefont {Reithmaier}}, \bibinfo {author} {\bibfnamefont
  {G.}~\bibnamefont {S{\c e}k}}, \ and\ \bibinfo {author} {\bibfnamefont
  {M.}~\bibnamefont {Benyoucef}},\ }\href@noop {} {\bibfield  {journal}
  {\bibinfo  {journal} {Appl. Phys. Lett.}\ }\textbf {\bibinfo {volume}
  {118}},\ \bibinfo {pages} {221101} (\bibinfo {year} {2021})}\BibitemShut
  {NoStop}%
\bibitem [{\citenamefont {Anderson}\ \emph {et~al.}(2021)\citenamefont
  {Anderson}, \citenamefont {M{\"u}ller}, \citenamefont {Skiba-Szymanska},
  \citenamefont {Krysa}, \citenamefont {Stevenson}, \citenamefont {Heffernan},
  \citenamefont {Ritchie},\ and\ \citenamefont {Shields}}]{Anderson_APL2021}%
  \BibitemOpen
  \bibfield  {author} {\bibinfo {author} {\bibfnamefont {M.}~\bibnamefont
  {Anderson}}, \bibinfo {author} {\bibfnamefont {T.}~\bibnamefont
  {M{\"u}ller}}, \bibinfo {author} {\bibfnamefont {J.}~\bibnamefont
  {Skiba-Szymanska}}, \bibinfo {author} {\bibfnamefont {J.~H.}\ \bibnamefont
  {Krysa}}, \bibinfo {author} {\bibfnamefont {R.~M.}\ \bibnamefont
  {Stevenson}}, \bibinfo {author} {\bibfnamefont {J.}~\bibnamefont
  {Heffernan}}, \bibinfo {author} {\bibfnamefont {D.~A.}\ \bibnamefont
  {Ritchie}}, \ and\ \bibinfo {author} {\bibfnamefont {A.~J.}\ \bibnamefont
  {Shields}},\ }\href@noop {} {\bibfield  {journal} {\bibinfo  {journal} {Appl.
  Phys. Lett.}\ }\textbf {\bibinfo {volume} {118}},\ \bibinfo {pages} {014003}
  (\bibinfo {year} {2021})}\BibitemShut {NoStop}%
\bibitem [{\citenamefont {Holewa}\ \emph {et~al.}(2022)\citenamefont {Holewa},
  \citenamefont {Sakanas}, \citenamefont {G{\"u}r}, \citenamefont
  {Mrowi{\'n}ski}, \citenamefont {Huck}, \citenamefont {Bi-Ying~Wang},
  \citenamefont {Yvind}, \citenamefont {Gregersen}, \citenamefont {Syperek},\
  and\ \citenamefont {Semenova}}]{Holewa_ACSP2022}%
  \BibitemOpen
  \bibfield  {author} {\bibinfo {author} {\bibfnamefont {P.}~\bibnamefont
  {Holewa}}, \bibinfo {author} {\bibfnamefont {A.}~\bibnamefont {Sakanas}},
  \bibinfo {author} {\bibfnamefont {U.~M.}\ \bibnamefont {G{\"u}r}}, \bibinfo
  {author} {\bibfnamefont {P.}~\bibnamefont {Mrowi{\'n}ski}}, \bibinfo {author}
  {\bibfnamefont {A.}~\bibnamefont {Huck}}, \bibinfo {author} {\bibfnamefont
  {A.~M.}\ \bibnamefont {Bi-Ying~Wang}}, \bibinfo {author} {\bibfnamefont
  {K.}~\bibnamefont {Yvind}}, \bibinfo {author} {\bibfnamefont
  {N.}~\bibnamefont {Gregersen}}, \bibinfo {author} {\bibfnamefont
  {M.}~\bibnamefont {Syperek}}, \ and\ \bibinfo {author} {\bibfnamefont
  {E.}~\bibnamefont {Semenova}},\ }\href@noop {} {\bibfield  {journal}
  {\bibinfo  {journal} {ACS Photon.}\ }\textbf {\bibinfo {volume} {9}},\
  \bibinfo {pages} {2273} (\bibinfo {year} {2022})}\BibitemShut {NoStop}%
\bibitem [{\citenamefont {Paul}\ \emph {et~al.}(2017)\citenamefont {Paul},
  \citenamefont {Olbrich}, \citenamefont {Hoschele}, \citenamefont {Schreier},
  \citenamefont {Kettler}, \citenamefont {Portalupi}, \citenamefont {Jetter},\
  and\ \citenamefont {Michler}}]{Paul_APL2017}%
  \BibitemOpen
  \bibfield  {author} {\bibinfo {author} {\bibfnamefont {M.}~\bibnamefont
  {Paul}}, \bibinfo {author} {\bibfnamefont {F.}~\bibnamefont {Olbrich}},
  \bibinfo {author} {\bibfnamefont {J.}~\bibnamefont {Hoschele}}, \bibinfo
  {author} {\bibfnamefont {S.}~\bibnamefont {Schreier}}, \bibinfo {author}
  {\bibfnamefont {J.}~\bibnamefont {Kettler}}, \bibinfo {author} {\bibfnamefont
  {S.~L.}\ \bibnamefont {Portalupi}}, \bibinfo {author} {\bibfnamefont
  {M.}~\bibnamefont {Jetter}}, \ and\ \bibinfo {author} {\bibfnamefont
  {P.}~\bibnamefont {Michler}},\ }\href@noop {} {\bibfield  {journal} {\bibinfo
   {journal} {Appl. Phys. Lett.}\ }\textbf {\bibinfo {volume} {111}},\ \bibinfo
  {pages} {033102} (\bibinfo {year} {2017})}\BibitemShut {NoStop}%
\bibitem [{\citenamefont {Carmesin}\ \emph {et~al.}(2018)\citenamefont
  {Carmesin}, \citenamefont {Olbrich}, \citenamefont {Mehrtens}, \citenamefont
  {Florian}, \citenamefont {Michael}, \citenamefont {Schreier}, \citenamefont
  {Nawrath}, \citenamefont {Paul}, \citenamefont {Hoschele}, \citenamefont
  {Gerken}, \citenamefont {Kettler}, \citenamefont {Portalupi}, \citenamefont
  {Jetter}, \citenamefont {Michler}, \citenamefont {Rosenauer},\ and\
  \citenamefont {Jahnke}}]{Carmesin_2018}%
  \BibitemOpen
  \bibfield  {author} {\bibinfo {author} {\bibfnamefont {C.}~\bibnamefont
  {Carmesin}}, \bibinfo {author} {\bibfnamefont {F.}~\bibnamefont {Olbrich}},
  \bibinfo {author} {\bibfnamefont {T.}~\bibnamefont {Mehrtens}}, \bibinfo
  {author} {\bibfnamefont {M.}~\bibnamefont {Florian}}, \bibinfo {author}
  {\bibfnamefont {S.}~\bibnamefont {Michael}}, \bibinfo {author} {\bibfnamefont
  {S.}~\bibnamefont {Schreier}}, \bibinfo {author} {\bibfnamefont
  {C.}~\bibnamefont {Nawrath}}, \bibinfo {author} {\bibfnamefont
  {M.}~\bibnamefont {Paul}}, \bibinfo {author} {\bibfnamefont {J.}~\bibnamefont
  {Hoschele}}, \bibinfo {author} {\bibfnamefont {B.}~\bibnamefont {Gerken}},
  \bibinfo {author} {\bibfnamefont {J.}~\bibnamefont {Kettler}}, \bibinfo
  {author} {\bibfnamefont {S.}~\bibnamefont {Portalupi}}, \bibinfo {author}
  {\bibfnamefont {M.}~\bibnamefont {Jetter}}, \bibinfo {author} {\bibfnamefont
  {P.}~\bibnamefont {Michler}}, \bibinfo {author} {\bibfnamefont
  {A.}~\bibnamefont {Rosenauer}}, \ and\ \bibinfo {author} {\bibfnamefont
  {F.}~\bibnamefont {Jahnke}},\ }\href@noop {} {\bibfield  {journal} {\bibinfo
  {journal} {Phys. Rev. B}\ }\textbf {\bibinfo {volume} {98}},\ \bibinfo
  {pages} {125407} (\bibinfo {year} {2018})}\BibitemShut {NoStop}%
\bibitem [{\citenamefont {Nawrath}\ \emph {et~al.}(2019)\citenamefont
  {Nawrath}, \citenamefont {Olbrich}, \citenamefont {Paul}, \citenamefont
  {Portalupi}, \citenamefont {Jetter},\ and\ \citenamefont
  {Michler}}]{Nawrath_APL2019}%
  \BibitemOpen
  \bibfield  {author} {\bibinfo {author} {\bibfnamefont {C.}~\bibnamefont
  {Nawrath}}, \bibinfo {author} {\bibfnamefont {F.}~\bibnamefont {Olbrich}},
  \bibinfo {author} {\bibfnamefont {M.}~\bibnamefont {Paul}}, \bibinfo {author}
  {\bibfnamefont {S.~L.}\ \bibnamefont {Portalupi}}, \bibinfo {author}
  {\bibfnamefont {M.}~\bibnamefont {Jetter}}, \ and\ \bibinfo {author}
  {\bibfnamefont {P.}~\bibnamefont {Michler}},\ }\href@noop {} {\bibfield
  {journal} {\bibinfo  {journal} {Appl. Phys. Lett.}\ }\textbf {\bibinfo
  {volume} {115}},\ \bibinfo {pages} {023103} (\bibinfo {year}
  {2019})}\BibitemShut {NoStop}%
\bibitem [{\citenamefont {Bauer}\ \emph {et~al.}(2021)\citenamefont {Bauer},
  \citenamefont {Wang}, \citenamefont {Hoppe}, \citenamefont {Nawrath},
  \citenamefont {Fischer}, \citenamefont {Witz}, \citenamefont {Kasche},
  \citenamefont {Schweikert}, \citenamefont {Jetter}, \citenamefont
  {Portalupi}, \citenamefont {Berroth},\ and\ \citenamefont
  {Michler}}]{Bauer_APL2021}%
  \BibitemOpen
  \bibfield  {author} {\bibinfo {author} {\bibfnamefont {S.}~\bibnamefont
  {Bauer}}, \bibinfo {author} {\bibfnamefont {D.}~\bibnamefont {Wang}},
  \bibinfo {author} {\bibfnamefont {N.}~\bibnamefont {Hoppe}}, \bibinfo
  {author} {\bibfnamefont {C.}~\bibnamefont {Nawrath}}, \bibinfo {author}
  {\bibfnamefont {J.}~\bibnamefont {Fischer}}, \bibinfo {author} {\bibfnamefont
  {N.}~\bibnamefont {Witz}}, \bibinfo {author} {\bibfnamefont {M.}~\bibnamefont
  {Kasche}}, \bibinfo {author} {\bibfnamefont {C.}~\bibnamefont {Schweikert}},
  \bibinfo {author} {\bibfnamefont {M.}~\bibnamefont {Jetter}}, \bibinfo
  {author} {\bibfnamefont {S.~L.}\ \bibnamefont {Portalupi}}, \bibinfo {author}
  {\bibfnamefont {M.}~\bibnamefont {Berroth}}, \ and\ \bibinfo {author}
  {\bibfnamefont {P.}~\bibnamefont {Michler}},\ }\href@noop {} {\bibfield
  {journal} {\bibinfo  {journal} {Appl. Phys. Lett.}\ }\textbf {\bibinfo
  {volume} {119}},\ \bibinfo {pages} {211101} (\bibinfo {year}
  {2021})}\BibitemShut {NoStop}%
\bibitem [{\citenamefont {Nawrath}\ \emph {et~al.}(2021)\citenamefont
  {Nawrath}, \citenamefont {Vural}, \citenamefont {Fischer}, \citenamefont
  {Schaber}, \citenamefont {Portalupi}, \citenamefont {Jetter},\ and\
  \citenamefont {Michler}}]{Nawrath_APL2021}%
  \BibitemOpen
  \bibfield  {author} {\bibinfo {author} {\bibfnamefont {C.}~\bibnamefont
  {Nawrath}}, \bibinfo {author} {\bibfnamefont {H.}~\bibnamefont {Vural}},
  \bibinfo {author} {\bibfnamefont {J.}~\bibnamefont {Fischer}}, \bibinfo
  {author} {\bibfnamefont {R.}~\bibnamefont {Schaber}}, \bibinfo {author}
  {\bibfnamefont {S.~L.}\ \bibnamefont {Portalupi}}, \bibinfo {author}
  {\bibfnamefont {M.}~\bibnamefont {Jetter}}, \ and\ \bibinfo {author}
  {\bibfnamefont {P.}~\bibnamefont {Michler}},\ }\href@noop {} {\bibfield
  {journal} {\bibinfo  {journal} {Appl. Phys. Lett.}\ }\textbf {\bibinfo
  {volume} {118}},\ \bibinfo {pages} {244002} (\bibinfo {year}
  {2021})}\BibitemShut {NoStop}%
\bibitem [{\citenamefont {Zeuner}\ \emph {et~al.}(2021)\citenamefont {Zeuner},
  \citenamefont {J{\"o}ns}, \citenamefont {Schweickert}, \citenamefont
  {Hedlund}, \citenamefont {Lobato}, \citenamefont {Lettner}, \citenamefont
  {Wang}, \citenamefont {Gyger}, \citenamefont {Sch{\"o}ll}, \citenamefont
  {Steinhauer}, \citenamefont {Hammar},\ and\ \citenamefont
  {Zwiller}}]{Zeuner_ACSP2021}%
  \BibitemOpen
  \bibfield  {author} {\bibinfo {author} {\bibfnamefont {K.}~\bibnamefont
  {Zeuner}}, \bibinfo {author} {\bibfnamefont {K.~D.}\ \bibnamefont
  {J{\"o}ns}}, \bibinfo {author} {\bibfnamefont {L.}~\bibnamefont
  {Schweickert}}, \bibinfo {author} {\bibfnamefont {C.~R.}\ \bibnamefont
  {Hedlund}}, \bibinfo {author} {\bibfnamefont {C.~N.}\ \bibnamefont {Lobato}},
  \bibinfo {author} {\bibfnamefont {T.}~\bibnamefont {Lettner}}, \bibinfo
  {author} {\bibfnamefont {K.}~\bibnamefont {Wang}}, \bibinfo {author}
  {\bibfnamefont {S.}~\bibnamefont {Gyger}}, \bibinfo {author} {\bibfnamefont
  {E.}~\bibnamefont {Sch{\"o}ll}}, \bibinfo {author} {\bibfnamefont
  {S.}~\bibnamefont {Steinhauer}}, \bibinfo {author} {\bibfnamefont
  {M.}~\bibnamefont {Hammar}}, \ and\ \bibinfo {author} {\bibfnamefont
  {V.}~\bibnamefont {Zwiller}},\ }\href@noop {} {\bibfield  {journal} {\bibinfo
   {journal} {ACS Photon.}\ }\textbf {\bibinfo {volume} {8}},\ \bibinfo {pages}
  {2337} (\bibinfo {year} {2021})}\BibitemShut {NoStop}%
\bibitem [{\citenamefont {Łukasz Dusanowski}\ \emph
  {et~al.}(2022)\citenamefont {Łukasz Dusanowski}, \citenamefont {Nawrath},
  \citenamefont {Portalupi}, \citenamefont {Jetter}, \citenamefont {Huber},
  \citenamefont {Klembt}, \citenamefont {Michler},\ and\ \citenamefont
  {H{\"o}fling}}]{Dusanowski_NC2022}%
  \BibitemOpen
  \bibfield  {author} {\bibinfo {author} {\bibnamefont {Łukasz Dusanowski}},
  \bibinfo {author} {\bibfnamefont {C.}~\bibnamefont {Nawrath}}, \bibinfo
  {author} {\bibfnamefont {S.~L.}\ \bibnamefont {Portalupi}}, \bibinfo {author}
  {\bibfnamefont {M.}~\bibnamefont {Jetter}}, \bibinfo {author} {\bibfnamefont
  {T.}~\bibnamefont {Huber}}, \bibinfo {author} {\bibfnamefont
  {S.}~\bibnamefont {Klembt}}, \bibinfo {author} {\bibfnamefont
  {P.}~\bibnamefont {Michler}}, \ and\ \bibinfo {author} {\bibfnamefont
  {S.}~\bibnamefont {H{\"o}fling}},\ }\href@noop {} {\bibfield  {journal}
  {\bibinfo  {journal} {Nat. Commun.}\ }\textbf {\bibinfo {volume} {13}},\
  \bibinfo {pages} {748} (\bibinfo {year} {2022})}\BibitemShut {NoStop}%
\bibitem [{\citenamefont {Takemoto}\ \emph {et~al.}(2015)\citenamefont
  {Takemoto}, \citenamefont {Nambu}, \citenamefont {Miyazawa}, \citenamefont
  {Sakuma}, \citenamefont {Yamamoto}, \citenamefont {Yorozu},\ and\
  \citenamefont {Arakawa}}]{Takemoto_SR2015}%
  \BibitemOpen
  \bibfield  {author} {\bibinfo {author} {\bibfnamefont {K.}~\bibnamefont
  {Takemoto}}, \bibinfo {author} {\bibfnamefont {Y.}~\bibnamefont {Nambu}},
  \bibinfo {author} {\bibfnamefont {T.}~\bibnamefont {Miyazawa}}, \bibinfo
  {author} {\bibfnamefont {Y.}~\bibnamefont {Sakuma}}, \bibinfo {author}
  {\bibfnamefont {T.}~\bibnamefont {Yamamoto}}, \bibinfo {author}
  {\bibfnamefont {S.}~\bibnamefont {Yorozu}}, \ and\ \bibinfo {author}
  {\bibfnamefont {Y.}~\bibnamefont {Arakawa}},\ }\href@noop {} {\bibfield
  {journal} {\bibinfo  {journal} {Sci. Rep.}\ }\textbf {\bibinfo {volume}
  {5}},\ \bibinfo {pages} {14383} (\bibinfo {year} {2015})}\BibitemShut
  {NoStop}%
\bibitem [{\citenamefont {Dalacu}\ \emph {et~al.}(2011)\citenamefont {Dalacu},
  \citenamefont {Mnaymneh}, \citenamefont {Wu}, \citenamefont {Lapointe},
  \citenamefont {Aers}, \citenamefont {Poole},\ and\ \citenamefont
  {Williams}}]{Dalacu_APL2011}%
  \BibitemOpen
  \bibfield  {author} {\bibinfo {author} {\bibfnamefont {D.}~\bibnamefont
  {Dalacu}}, \bibinfo {author} {\bibfnamefont {K.}~\bibnamefont {Mnaymneh}},
  \bibinfo {author} {\bibfnamefont {X.}~\bibnamefont {Wu}}, \bibinfo {author}
  {\bibfnamefont {J.}~\bibnamefont {Lapointe}}, \bibinfo {author}
  {\bibfnamefont {G.~C.}\ \bibnamefont {Aers}}, \bibinfo {author}
  {\bibfnamefont {P.~J.}\ \bibnamefont {Poole}}, \ and\ \bibinfo {author}
  {\bibfnamefont {R.~L.}\ \bibnamefont {Williams}},\ }\href@noop {} {\bibfield
  {journal} {\bibinfo  {journal} {Appl. Phys. Lett.}\ }\textbf {\bibinfo
  {volume} {98}},\ \bibinfo {pages} {251101} (\bibinfo {year}
  {2011})}\BibitemShut {NoStop}%
\bibitem [{\citenamefont {Dalacu}\ \emph {et~al.}(2019)\citenamefont {Dalacu},
  \citenamefont {Poole},\ and\ \citenamefont {Williams}}]{Dalacu_NT2019}%
  \BibitemOpen
  \bibfield  {author} {\bibinfo {author} {\bibfnamefont {D.}~\bibnamefont
  {Dalacu}}, \bibinfo {author} {\bibfnamefont {P.~J.}\ \bibnamefont {Poole}}, \
  and\ \bibinfo {author} {\bibfnamefont {R.~L.}\ \bibnamefont {Williams}},\
  }\href@noop {} {\bibfield  {journal} {\bibinfo  {journal} {Nanotechnol.}\
  }\textbf {\bibinfo {volume} {30}},\ \bibinfo {pages} {232001} (\bibinfo
  {year} {2019})}\BibitemShut {NoStop}%
\bibitem [{\citenamefont {Laferri{\`e}re}\ \emph {et~al.}(2022)\citenamefont
  {Laferri{\`e}re}, \citenamefont {Yeung}, \citenamefont {Miron}, \citenamefont
  {Northeast}, \citenamefont {Haffouz}, \citenamefont {Lapointe}, \citenamefont
  {Korkusinski}, \citenamefont {Poole}, \citenamefont {Williams},\ and\
  \citenamefont {Dalacu}}]{Laferriere_SR2022}%
  \BibitemOpen
  \bibfield  {author} {\bibinfo {author} {\bibfnamefont {P.}~\bibnamefont
  {Laferri{\`e}re}}, \bibinfo {author} {\bibfnamefont {E.}~\bibnamefont
  {Yeung}}, \bibinfo {author} {\bibfnamefont {I.}~\bibnamefont {Miron}},
  \bibinfo {author} {\bibfnamefont {D.~B.}\ \bibnamefont {Northeast}}, \bibinfo
  {author} {\bibfnamefont {S.}~\bibnamefont {Haffouz}}, \bibinfo {author}
  {\bibfnamefont {J.}~\bibnamefont {Lapointe}}, \bibinfo {author}
  {\bibfnamefont {M.}~\bibnamefont {Korkusinski}}, \bibinfo {author}
  {\bibfnamefont {P.~J.}\ \bibnamefont {Poole}}, \bibinfo {author}
  {\bibfnamefont {R.~L.}\ \bibnamefont {Williams}}, \ and\ \bibinfo {author}
  {\bibfnamefont {D.}~\bibnamefont {Dalacu}},\ }\href@noop {} {\bibfield
  {journal} {\bibinfo  {journal} {Sci. Rep.}\ }\textbf {\bibinfo {volume}
  {12}},\ \bibinfo {pages} {6376} (\bibinfo {year} {2022})}\BibitemShut
  {NoStop}%
\bibitem [{\citenamefont {Dalacu}\ \emph {et~al.}(2012)\citenamefont {Dalacu},
  \citenamefont {Mnaymneh}, \citenamefont {Lapointe}, \citenamefont {Wu},
  \citenamefont {Poole}, \citenamefont {Bulgarini}, \citenamefont {Zwiller},\
  and\ \citenamefont {Reimer}}]{Dalacu_NL2012}%
  \BibitemOpen
  \bibfield  {author} {\bibinfo {author} {\bibfnamefont {D.}~\bibnamefont
  {Dalacu}}, \bibinfo {author} {\bibfnamefont {K.}~\bibnamefont {Mnaymneh}},
  \bibinfo {author} {\bibfnamefont {J.}~\bibnamefont {Lapointe}}, \bibinfo
  {author} {\bibfnamefont {X.}~\bibnamefont {Wu}}, \bibinfo {author}
  {\bibfnamefont {P.}~\bibnamefont {Poole}}, \bibinfo {author} {\bibfnamefont
  {G.}~\bibnamefont {Bulgarini}}, \bibinfo {author} {\bibfnamefont
  {V.}~\bibnamefont {Zwiller}}, \ and\ \bibinfo {author} {\bibfnamefont
  {M.}~\bibnamefont {Reimer}},\ }\href@noop {} {\bibfield  {journal} {\bibinfo
  {journal} {Nano Lett.}\ }\textbf {\bibinfo {volume} {12}},\ \bibinfo {pages}
  {5919} (\bibinfo {year} {2012})}\BibitemShut {NoStop}%
\bibitem [{\citenamefont {Laferri{\`e}re}\ \emph {et~al.}(2023)\citenamefont
  {Laferri{\`e}re}, \citenamefont {Haffouz}, \citenamefont {Northeast},
  \citenamefont {Poole}, \citenamefont {Williams},\ and\ \citenamefont
  {Dalacu}}]{Laferriere_NL2023}%
  \BibitemOpen
  \bibfield  {author} {\bibinfo {author} {\bibfnamefont {P.}~\bibnamefont
  {Laferri{\`e}re}}, \bibinfo {author} {\bibfnamefont {S.}~\bibnamefont
  {Haffouz}}, \bibinfo {author} {\bibfnamefont {D.~B.}\ \bibnamefont
  {Northeast}}, \bibinfo {author} {\bibfnamefont {P.~J.}\ \bibnamefont
  {Poole}}, \bibinfo {author} {\bibfnamefont {R.~L.}\ \bibnamefont {Williams}},
  \ and\ \bibinfo {author} {\bibfnamefont {D.}~\bibnamefont {Dalacu}},\
  }\href@noop {} {\bibfield  {journal} {\bibinfo  {journal} {Nano Lett.}\
  }\textbf {\bibinfo {volume} {23}},\ \bibinfo {pages} {962} (\bibinfo {year}
  {2023})}\BibitemShut {NoStop}%
\bibitem [{\citenamefont {Chauvin}\ \emph {et~al.}(2012)\citenamefont
  {Chauvin}, \citenamefont {Alouane}, \citenamefont {Anufriev}, \citenamefont
  {Khmissi}, \citenamefont {Naji}, \citenamefont {Patriarche}, \citenamefont
  {Bru-Chevallier},\ and\ \citenamefont {Gendry}}]{Chauvin_APL2012}%
  \BibitemOpen
  \bibfield  {author} {\bibinfo {author} {\bibfnamefont {N.}~\bibnamefont
  {Chauvin}}, \bibinfo {author} {\bibfnamefont {M.~H.~H.}\ \bibnamefont
  {Alouane}}, \bibinfo {author} {\bibfnamefont {R.}~\bibnamefont {Anufriev}},
  \bibinfo {author} {\bibfnamefont {H.}~\bibnamefont {Khmissi}}, \bibinfo
  {author} {\bibfnamefont {K.}~\bibnamefont {Naji}}, \bibinfo {author}
  {\bibfnamefont {G.}~\bibnamefont {Patriarche}}, \bibinfo {author}
  {\bibfnamefont {C.}~\bibnamefont {Bru-Chevallier}}, \ and\ \bibinfo {author}
  {\bibfnamefont {M.}~\bibnamefont {Gendry}},\ }\href@noop {} {\bibfield
  {journal} {\bibinfo  {journal} {Appl. Phys. Lett.}\ }\textbf {\bibinfo
  {volume} {100}},\ \bibinfo {pages} {011906} (\bibinfo {year}
  {2012})}\BibitemShut {NoStop}%
\bibitem [{\citenamefont {Haffouz}\ \emph {et~al.}(2020)\citenamefont
  {Haffouz}, \citenamefont {Poole}, \citenamefont {Jin}, \citenamefont {Wu},
  \citenamefont {Mnaymneh}, \citenamefont {Dalacu},\ and\ \citenamefont
  {Williams}}]{Haffouz_APL2020}%
  \BibitemOpen
  \bibfield  {author} {\bibinfo {author} {\bibfnamefont {S.}~\bibnamefont
  {Haffouz}}, \bibinfo {author} {\bibfnamefont {P.~J.}\ \bibnamefont {Poole}},
  \bibinfo {author} {\bibfnamefont {J.}~\bibnamefont {Jin}}, \bibinfo {author}
  {\bibfnamefont {L.}~\bibnamefont {Wu}, \bibfnamefont {X.~Ginet}}, \bibinfo
  {author} {\bibfnamefont {K.}~\bibnamefont {Mnaymneh}}, \bibinfo {author}
  {\bibfnamefont {D.}~\bibnamefont {Dalacu}}, \ and\ \bibinfo {author}
  {\bibfnamefont {R.~L.}\ \bibnamefont {Williams}},\ }\href@noop {} {\bibfield
  {journal} {\bibinfo  {journal} {Appl. Phys. Lett.}\ }\textbf {\bibinfo
  {volume} {117}},\ \bibinfo {pages} {113102} (\bibinfo {year}
  {2020})}\BibitemShut {NoStop}%
\bibitem [{\citenamefont {Haffouz}\ \emph {et~al.}(2018)\citenamefont
  {Haffouz}, \citenamefont {Zeuner}, \citenamefont {Dalacu}, \citenamefont
  {Poole}, \citenamefont {Lapointe}, \citenamefont {Poitras}, \citenamefont
  {Mnaymneh}, \citenamefont {Wu}, \citenamefont {Couillard}, \citenamefont
  {Korkusinski}, \citenamefont {Sch{\"o}ll}, \citenamefont {J{\"o}ns},
  \citenamefont {Zwiller},\ and\ \citenamefont {Williams}}]{Haffouz_NL2018}%
  \BibitemOpen
  \bibfield  {author} {\bibinfo {author} {\bibfnamefont {S.}~\bibnamefont
  {Haffouz}}, \bibinfo {author} {\bibfnamefont {K.~D.}\ \bibnamefont {Zeuner}},
  \bibinfo {author} {\bibfnamefont {D.}~\bibnamefont {Dalacu}}, \bibinfo
  {author} {\bibfnamefont {P.~J.}\ \bibnamefont {Poole}}, \bibinfo {author}
  {\bibfnamefont {J.}~\bibnamefont {Lapointe}}, \bibinfo {author}
  {\bibfnamefont {D.}~\bibnamefont {Poitras}}, \bibinfo {author} {\bibfnamefont
  {K.}~\bibnamefont {Mnaymneh}}, \bibinfo {author} {\bibfnamefont
  {X.}~\bibnamefont {Wu}}, \bibinfo {author} {\bibfnamefont {M.}~\bibnamefont
  {Couillard}}, \bibinfo {author} {\bibfnamefont {M.}~\bibnamefont
  {Korkusinski}}, \bibinfo {author} {\bibfnamefont {E.}~\bibnamefont
  {Sch{\"o}ll}}, \bibinfo {author} {\bibfnamefont {K.~D.}\ \bibnamefont
  {J{\"o}ns}}, \bibinfo {author} {\bibfnamefont {V.}~\bibnamefont {Zwiller}}, \
  and\ \bibinfo {author} {\bibfnamefont {R.~L.}\ \bibnamefont {Williams}},\
  }\href@noop {} {\bibfield  {journal} {\bibinfo  {journal} {Nano Lett.}\
  }\textbf {\bibinfo {volume} {18}},\ \bibinfo {pages} {3047} (\bibinfo {year}
  {2018})}\BibitemShut {NoStop}%
\bibitem [{XX()}]{XX}%
  \BibitemOpen
  \href@noop {} {\bibinfo  {journal} {We note that the quoted $XX$ decay time
  is for $\lambda = 950$nm emitters and it is likely that this decay time is
  wavelength dependent.}\ }\BibitemShut {NoStop}%
\bibitem [{\citenamefont {Laferrière}\ \emph {et~al.}(2020)\citenamefont
  {Laferrière}, \citenamefont {Yeung}, \citenamefont {Giner}, \citenamefont
  {Haffouz}, \citenamefont {Lapointe}, \citenamefont {Aers}, \citenamefont
  {Poole}, \citenamefont {Williams},\ and\ \citenamefont
  {Dalacu}}]{Laferriere_NL2020}%
  \BibitemOpen
\bibfield  {journal} {  }\bibfield  {author} {\bibinfo {author} {\bibfnamefont
  {P.}~\bibnamefont {Laferrière}}, \bibinfo {author} {\bibfnamefont
  {E.}~\bibnamefont {Yeung}}, \bibinfo {author} {\bibfnamefont
  {L.}~\bibnamefont {Giner}}, \bibinfo {author} {\bibfnamefont
  {S.}~\bibnamefont {Haffouz}}, \bibinfo {author} {\bibfnamefont
  {J.}~\bibnamefont {Lapointe}}, \bibinfo {author} {\bibfnamefont {G.~C.}\
  \bibnamefont {Aers}}, \bibinfo {author} {\bibfnamefont {P.~J.}\ \bibnamefont
  {Poole}}, \bibinfo {author} {\bibfnamefont {R.~L.}\ \bibnamefont {Williams}},
  \ and\ \bibinfo {author} {\bibfnamefont {D.}~\bibnamefont {Dalacu}},\ }\href
  {\doibase 10.1021/acs.nanolett.0c00607} {\bibfield  {journal} {\bibinfo
  {journal} {Nano Lett.}\ }\textbf {\bibinfo {volume} {20}},\ \bibinfo {pages}
  {3688} (\bibinfo {year} {2020})}\BibitemShut {NoStop}%
\bibitem [{\citenamefont {Mnaymneh}\ \emph {et~al.}(2020)\citenamefont
  {Mnaymneh}, \citenamefont {Dalacu}, \citenamefont {McKee}, \citenamefont
  {Lapointe}, \citenamefont {Haffouz}, \citenamefont {Weber}, \citenamefont
  {Northeast}, \citenamefont {Poole}, \citenamefont {Aers},\ and\ \citenamefont
  {Williams}}]{Mnaymneh_AQT2020}%
  \BibitemOpen
  \bibfield  {author} {\bibinfo {author} {\bibfnamefont {K.}~\bibnamefont
  {Mnaymneh}}, \bibinfo {author} {\bibfnamefont {D.}~\bibnamefont {Dalacu}},
  \bibinfo {author} {\bibfnamefont {J.}~\bibnamefont {McKee}}, \bibinfo
  {author} {\bibfnamefont {J.}~\bibnamefont {Lapointe}}, \bibinfo {author}
  {\bibfnamefont {S.}~\bibnamefont {Haffouz}}, \bibinfo {author} {\bibfnamefont
  {J.~F.}\ \bibnamefont {Weber}}, \bibinfo {author} {\bibfnamefont {D.~B.}\
  \bibnamefont {Northeast}}, \bibinfo {author} {\bibfnamefont {P.~J.}\
  \bibnamefont {Poole}}, \bibinfo {author} {\bibfnamefont {G.~C.}\ \bibnamefont
  {Aers}}, \ and\ \bibinfo {author} {\bibfnamefont {R.~L.}\ \bibnamefont
  {Williams}},\ }\href@noop {} {\bibfield  {journal} {\bibinfo  {journal} {Adv.
  Quant. Tech.}\ }\textbf {\bibinfo {volume} {3}},\ \bibinfo {pages} {1900021}
  (\bibinfo {year} {2020})}\BibitemShut {NoStop}%
\bibitem [{\citenamefont {Aichele}\ \emph {et~al.}(2004)\citenamefont
  {Aichele}, \citenamefont {Zwiller},\ and\ \citenamefont
  {Benson}}]{Aichele_NJP2004}%
  \BibitemOpen
  \bibfield  {author} {\bibinfo {author} {\bibfnamefont {T.}~\bibnamefont
  {Aichele}}, \bibinfo {author} {\bibfnamefont {V.}~\bibnamefont {Zwiller}}, \
  and\ \bibinfo {author} {\bibfnamefont {O.}~\bibnamefont {Benson}},\
  }\href@noop {} {\bibfield  {journal} {\bibinfo  {journal} {New J. Phys.}\
  }\textbf {\bibinfo {volume} {6}},\ \bibinfo {pages} {90} (\bibinfo {year}
  {2004})}\BibitemShut {NoStop}%
\bibitem [{\citenamefont {Santori}\ \emph {et~al.}(2004)\citenamefont
  {Santori}, \citenamefont {Fattal}, \citenamefont {Vu{\v c}kovi{\' c}},
  \citenamefont {Solomon},\ and\ \citenamefont {Yamamoto}}]{Santori_NJP2004}%
  \BibitemOpen
  \bibfield  {author} {\bibinfo {author} {\bibfnamefont {C.}~\bibnamefont
  {Santori}}, \bibinfo {author} {\bibfnamefont {D.}~\bibnamefont {Fattal}},
  \bibinfo {author} {\bibfnamefont {J.}~\bibnamefont {Vu{\v c}kovi{\' c}}},
  \bibinfo {author} {\bibfnamefont {G.~S.}\ \bibnamefont {Solomon}}, \ and\
  \bibinfo {author} {\bibfnamefont {Y.}~\bibnamefont {Yamamoto}},\ }\href@noop
  {} {\bibfield  {journal} {\bibinfo  {journal} {New. J. Phys.}\ }\textbf
  {\bibinfo {volume} {6}},\ \bibinfo {pages} {89} (\bibinfo {year}
  {2004})}\BibitemShut {NoStop}%
\bibitem [{\citenamefont {Dalacu}\ \emph {et~al.}(2021)\citenamefont {Dalacu},
  \citenamefont {Poole},\ and\ \citenamefont {Williams}}]{Dalacu_NANOM2021}%
  \BibitemOpen
  \bibfield  {author} {\bibinfo {author} {\bibfnamefont {D.}~\bibnamefont
  {Dalacu}}, \bibinfo {author} {\bibfnamefont {P.~J.}\ \bibnamefont {Poole}}, \
  and\ \bibinfo {author} {\bibfnamefont {R.~L.}\ \bibnamefont {Williams}},\
  }\href@noop {} {\bibfield  {journal} {\bibinfo  {journal} {Nanomaterials}\
  }\textbf {\bibinfo {volume} {11}},\ \bibinfo {pages} {1201} (\bibinfo {year}
  {2021})}\BibitemShut {NoStop}%
\end{thebibliography}%
\end{document}